# DIELECTRIC AND PYROELECTRIC PROPERTIES OF THICK FILM FERROMAGNETIC - PIEZOELECTRIC STRUCTURES

Y.K. Fetisov, A.A. Bush K.E. Kamentsev
*Moscow Institute of Radioengineering, Electronics and Automation, 117454 Moscow, Russia*

G. Srinivasan
*Physics Department, Oakland University, Rochester, Michigan 48309-4401*

## Abstract

Layered ferromagnetic-piezoelectric composites show mechanical strain mediated electromagnetic coupling. Here we discuss dielectric and piezoelectric properties of ferrite-lead zirconate titanate (PZT) and lanthanum manganite-PZT samples. Results of our investigations on dielectric and pyroelectric properties of multilayer ferromagnetic-piezoelectric are presented here. Lead zircinate-titanate $PbZr_xTi_{1-x}O3$ (PZT) was used for the piezoelectric phase in all the structures. The following materials were used for the ferromagnetic component: nickel-zinc ferrites $Ni_{0.9}Zn_{0.1}Fe_2O_4$ (NFO1) and $Ni_{0.8}Zn_{0.2}Fe_2O_4$ (NFO2), cobalt ferrite (CFO), lithium ferrite (LFO), lanthanum strontium manganite $La_{0.7}Sr_{0.3}MnO_3$ (LSM), and lanthanum-calcium manganite $La_{0.7}Ca_{0.3}MnO_3$ (LCM). The pyroelectric effect was studied by measuring the current J flowing through a closed loop containing the sample and an electrometer as the sample temperature T was slowly varied at the rate 0.1 K/s. Polarized PZT layers generate a pyroelectric current as the temperature changes. The main indicator of pyroelectric nature of the current is the sign reversal when the thermal cycle is switched from heating to cooling. Almost all of the multilayer structures showed a pyroelectric current, but the pyroelectric coefficient varied in a wide range. (i) For NFO1-PZT system the coefficient was in the range 0.01 - 10 nC/($cm^2$ K), depending on the temperature. (ii) CFO-PZT and LFO-PZT structures showed a large thermal current and a weak pyroelectric effect. (iii) Thermal currents, however, were absent in LCM-PZT within the temperature range from the room temperature to 400 K. (iv) In LSM-PZT, the thermal current exceeded the pyroelectric current. A model is proposed for an understanding of these results.

## I. Introduction

Multilayer composite structures consisting of alternate layers of ferromagnetic and piezoelectric materials are extensively studied in recent years because of a giant magnetoelectric effect discovered in these structures [1]. Magnetoelectric effect in the composite structures arises from "product-properties" of the constituting layers [2]. When the composite structure is placed in external magnetic field *H*, a deformation of the magnetic layer due to magnetostriction results in a deformation of the piezoelectric layer. The deformation of the piezoelectric layer results in its polarization due to piezoelectric effect. The value of polarization is connected with the value of applied magnetic field by the relation *P*=α*H*, where α is the second rank magnetoelectric tensor [3]. Components of the tensor are proportional to the product of the magnetostriction constant $\lambda_{ij}$ of the ferromagnetic layer and piezoelectric constant $d_{jk}$ of the piezoelectric layer $\alpha_{ik} \sim \lambda_{ij} \ d_{jk}$. Polarization of the structure results in appearance of charges on its surfaces and creation of electric field *E* inside the structure.

In such structures, one usually measures alternative electric field produced due to magnetoelectric effect when the sample is placed in external alternative magnetic field. Value of the alternative electric field δ*E* in the piezoelectric layer is connected with value of the alternative magnetic field δ*H* by the relation δ*E* = $\alpha_E$·δ*H*, where $\alpha_E = \alpha/(\varepsilon_0 \cdot \varepsilon)$ is the magnetoelectric voltage coefficient, $\varepsilon_0$ is the dielectric constant, and ε is the dielectric permittivity.

The magnetoelectric effect with values of $\alpha_E$ =10-500 mV/cm Oe has been observed experimentally in multilayer ferrite-piezoelectric structures [1,4-7]. Calculations based on magnetostrictive and piezoelectric parameters of the two phases predict the possibility of even higher magnetoelectric coefficients [8]. Limitations on observed value of the magnetoelectric coefficient may be due to deterioration of magnetic and electrical properties of materials during the structures fabrication process. In particular, magnetoelectric coefficient depends strongly on the dielectric permeability and polarization of the piezoelectric layers, on the conductivity of piezoelectric and magnetostrictive layers. These properties can differ significantly from similar properties of volume materials used for the structure fabrication. Detailed investigations of dielectric properties of multilayer structures are required in order to understand the interdependence of the magnetoelectric and dielectric properties of multilayer structures.

Piezoelectric layers in multilayer magnetoelectric structures are usually fabricated of ferroelectric materials which posses both high piezoelectric and high pyroelectric effect. The pyroelectric effect [9] exhibits itself as a change in spontaneous polarization of the sample δ*P* caused by a change in the sample temperature δ*T*, where **γ** is the pyroelectric coefficient. The change in the sample polarization results in the appearance of charges on the sample surfaces and the creation of an electric field. The strength of the field δ*E* is related to the temperature change δ*T* by δ*E* = $\gamma/(\varepsilon \cdot \varepsilon_0) \cdot \delta T$. This slowly varying electrical field produces an electrical current in an external circuit and is measured in experiments. It will be shown here that studies of pyroelectric effect lead to information about quality of magnetoelectric structures and the strength of magnetoelectric (ME) effect in such structures.

We provide here results of our investigations of dielectric and pyroelectric properties of multilayer ferromagnetic-piezoelectric structures. Section 2 gives a short description of the fabrication of the structures. A ferroelectric lead zirconate titanate (PZT) composition with high piezoelectric effect at room temperature was used as the piezoelectric phase. Nickel ferrite, lithium ferrite, cobalt ferrite, or manganites with high magnetostariction was used as magnetic phase.

Experimental set up and methods of investigations of electrical and pyroelectric characteristics in multilayer structures are described in Section 3.

Data on temperature dependences of dielectric constant and resistivity of the structures are given in Section 4. The results are then discussed in the frame of multilayer structure model in Section 5. Data on temperature dependence of the pyroelectric constant for multilayers structures are summarized in Section 6. Results of investigations on magnetoelectric effect and its connection with dielectric parameters of the structures are discussed in Section 7.

## 2. Fabrication and parameters of multilayer structures

Multilayer structures consisting of alternate layers of piezoelectric and various ferromagnets were used in experiments. Lead zircinate-titanate (PZT) was used as piezoelectric phase in all the structures. The following materials were used as ferromagnets: nickel-zinc ferrites of the contents $Ni_{0.9}Zn_{0.1}Fe_2O_4$ (NFO1) and $Ni_{0.8}Zn_{0.2}Fe_2O_4$ (NFO2), cobalt ferrite $CoFe_2O_4$ (CFO), lithium ferrite $Li_{0.5}Fe_{2.5}O_4$ (LFO), lanthanum strontium manganite $La_{0.7}Sr_{0.3}MnO_3$ (LSM), and lanthanum-calcium manganite $La_{0.7}Ca_{0.3}MnO_3$ (LCM). Multilayer structures were synthesized from thick films fabricated by tape casting [10]. The ferrite and manganite powders necessary for the films were prepared using standard ceramic techniques including mixing of oxides and carbonates of corresponding metals followed by their sintering at high temperature. The oxides were ballmilled to obtain submicron-size powders. Commercially available PZT powder was used [11]. The process of thick film fabrication included the following steps: mixing of the powers with a solvent (ethyl alcohol) and a dispersant (Blown Menhaden fish oil) and ball milling for 24 h; second ball milling with a plasticizer (butyl benzyl phthalate) and a binder (polyvinyl butyral) for 24 h; casting the slurries thus obtained on silicon coated mylar sheets in a shape of 10-40 μm thick films using a mechanical tape caster; and heating the films at room temperature for 24 h followed by their separation from the mylar sheets. The tapes thus fabricated were arranged to obtain the desired structure, laminated under high pressure (3000-5000 psi) and temperature 400 K, and heated at 1000 K for binder evaporation. The final sintering was carried out at 1400-1500 K.

Thickness $d_f$ of magnetic layers and thickness $d_p$ of piezoelectric layers in range 10 -108 μm, number of PZT layers was $n = 8$-20, and the number of ferromagnetic layers was $n+1$. Samples studied, their constituents, number and thickness of layers, and areas of the structures are given in Table 1. Inside each group the structures are numbered according to volume for the two phases. Room temperature DC electrical resistivity $\rho$ and dielectric permeability $\varepsilon$ are given in Table 1.

Table 1. Parameters of multilayer structures

| N | Magnetic layer $d_f$ x $(n+1)$ | PZT layer $d_p$ x $n$ | S, mm$^2$ | ρ, $10^6$ Ω·cm | $\varepsilon_{ef}$ |
|---|---|---|---|---|---|
| NFO1– PZT | | | | | |
| 1 | 18 μm x 16 | 72 μm x 15 | 23 | 517 | 215 |
| 2 | 18 μm x 16 | 36 μm x 15 | 17 | 31 | 83 |
| 3 | 18 μm x 16 | 18 μm x 15 | 20 | 46 | 51 |
| 4 | 36 μm x 16 | 18 μm x 15 | 20 | 202 | 101 |
| 5 | 72 μm x 16 | 18 μm x 15 | 30 | 738 | 97 |
| NFO2– PZT | | | | | |
| 1 | 18 μm x 16 | 72μm x 15 | 37 | 105 | 159 |
| 2 | 18 μm x 16 | 36 μm x 15 | 30 | 228 | 95 |
| 3 | 18 μm x 16 | 18 μm x 15 | 19 | 289 | 87 |
| 4 | 36 μm x 16 | 18 μm x 15 | 28 | - | - |
| 5 | 72 μm x 16 | 18 μm x 15 | 42 | 241 | 62 |
| CFO – PZT | | | | | |
| 1 | 36 μm x 11 | 36 μm x 10 | 20 | 771 | 31.3 |
| 2 | 36 μm x 16 | 36 μm x 15 | 27 | 1227 | 51 |
| 3 | 36 μm x 21 | 36 μm x 20 | 32 | 278 | 104 |
| 4 | 36 μm x 26 | 36 μm x 25 | 25 | 230 | 162 |
| LFO – PZT | | | | | |
| 1 | 36 μm x 11 | 72 μm x 10 | 19 | 59 | 378 |
| 2 | 36 μm x 11 | 36 μm x 10 | 23 | 23 | 1175 |
| 3 | 72 μm x 11 | 36 μm x 10 | 25 | 13 | 2744 |
| 4 | 108 μm x 11 | 36 μm x 10 | 22 | 108 | 938 |
| LSM – PZT | | | | | |
| | 10 μm x 21 | 10 μm x 20 | 24 | 396 | 22 |
| LCM – PZT | | | | | |
| | 35 μm x 9 | 35 μm x 8 | 18 | 117 | 74 |

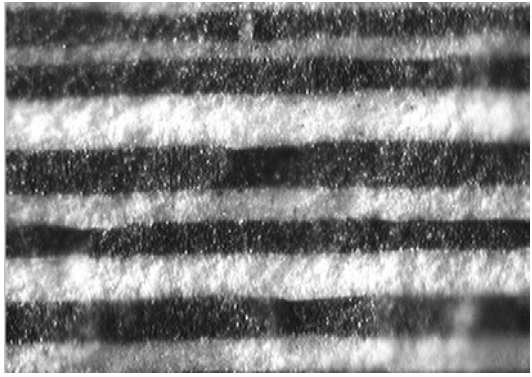

Fig.1. Fragment of cross-section of multilayer structure NFO1–PZT # 3.

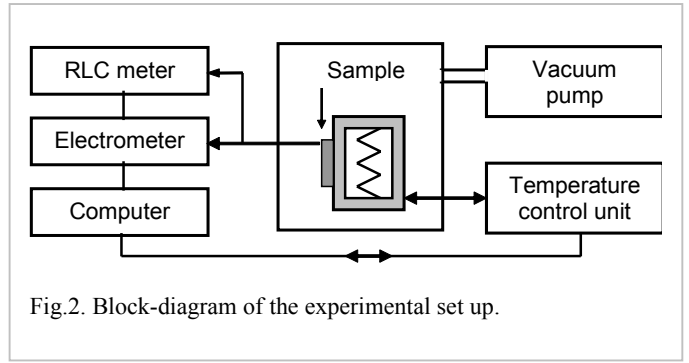

Fig.2. Block-diagram of the experimental set up.

The samples were then electrically poled. The sample surfaces were covered with a silver paint, it was placed in permanent electrical field of $E = 20$ kV/cm directed perpendicular to the sample plane and heated up to 420 K. As the sample was cooled to 300 K, the field strength $E$ was increased progressively to 50 kV/cm over duration of 30 min.

The surface morphology and cross section of the samples were examined with an optical microscope. Samples contained fine grains (1-5 $\mu$m) and some open pores. Figure 1 shows typical cross-section view of the NFO1-PZT #3 sample with expected nickel ferrite layer thickness of 18 $\mu$m and PZT layers thickness of 18 $\mu$m. The ferrite layers are of black color and PZT layers are of white color. One can see that, in this particular case, thickness of both ferrite and PZT layers differ significantly from the expected value, while thickness of each separate layer remains nearly constant along the sample length. This could result from non uniform pressure during lamination and/or an irregular shrinkage of the layers during the sintering process. Besides that, one can see separate violations of continuity of thin PZT layers. But most other samples showed uniform structures with the expected layer thickness.

Multilayers and powders prepared from these thick films were characterized using an X-ray diffractometer. The measurements were carried out using the Cu-K$_\alpha$ filtered radiation at a scan velocity of 0.5 deg/mm. The X-ray diffraction patterns of the powders contained two sets of well defined narrow peaks. The first set corresponded to the magnetic phase (NFO, CFO, LFO or manganites), while the second set was identified with a PZT phase. Main peaks of both sets were of nearly the same intensity. Lattice parameters of the phases calculated using the diffraction patterns were in good agreement with corresponding values for bulk materials [12]. The diffraction patterns obtained from the surfaces of the samples contained the same two sets of peaks. However, the intensity of PZT peaks was 10% of the nominal value due to shielding of internal PZT layers by external magnetic layer.

The results of optical and X-ray investigations led to conclude that: (i) layered structure is conserved in the samples during the sintering process; (ii) samples have the same crystal structures as individual phases, and (ii) no any new (impurity) phases are formed at the interfaces in the layered samples because of undesired diffusion.

## 3. Experimental set up and measurement technique

Figure 2 shows a block-diagram of the experimental set-up used for investigation of electrical characteristics of multilayer structures. The installation consisted of an electrically shielded vacuum cell, vacuum pump, heating elements, temperature sensor placed inside the vacuum cell, temperature control unit, digital $RLC$-meter, digital electrometer for current measurements, and a computer for data acquisition. The sample under test was placed inside the vacuum cell in close contact with heating elements. The pump provided pressure less than $\sim 10^{-3}$ mm of Hg. The temperature control unit allowed to maintain and change the temperature the 290-870 K range with an accuracy better than 1 K. Platinum wires were used to get signal out from the sample. The $RLC$-meter allowed measurements of capacitance and resistivity within the mentioned temperature range at $f = 100$ Hz, 1 kHz and 10 kHz. The electrometer had an input resistance higher than $10^{12}$ $\Omega$, and a characteristic time about of $\sim$3-30 ms and allowed measurements of current up to $2 \cdot 10^{-15}$ A.

In investigations on the pyroelectric effect, the current $J$ flowing trough a close circuit containing the sample and the electrometer was measured at slow variation of the sample temperature $T$. The sample temperature was increased nearly linearly at the rate $\partial T / \partial t = 0.1$ K/s during the sample heating. Then, the temperature was decreased nearly exponentially with characteristic time about of $\sim$1 hour during the sample cooling. Temperature dependence of the pyroelectric coefficient $\gamma$ for multilayer structures was then calculated using the data and the relation $\gamma = j/(dT/dt)$, where $j = J/S$ is the density of electrical current in the sample.

## 4. Dielectric properties

### 4.1. Temperature dependence

At the first stage, the temperature depeddence of electrical characteristics for all the structures listed in Table 1 were carried out. Figure 3 and Fig.4 show representative data on the effective dielectric permittivity $\varepsilon$, the specific resistance $\rho$, and the dielectric loss factor $tg\delta = 1/(2\pi f \varepsilon \varepsilon_0 \rho)$ vs. temperature for nickel ferrite and cobalt ferrite structures, respectively. The arrows on the figures correspond to the increase or decrease of the sample temperature during measurements. All the dependences were calculated using values of capacitance $C$ and resistance $R$ of the structures measured at the operating frequency $f$=1 kHz and taking into account the thicknesses and areas of the structures, as indicated in Table 1.

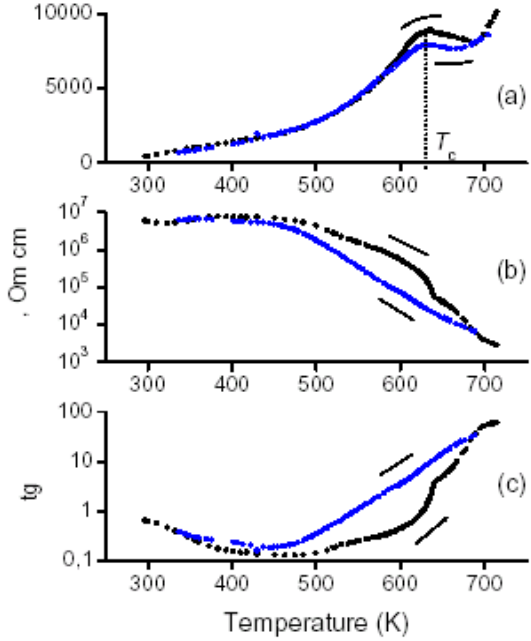

Fig. 3. Dependences of the dielectric permeability (a), specific resistance (b), and dielectric loss factor (c) vs. temperature for the NFO1-PZT #1 structure.

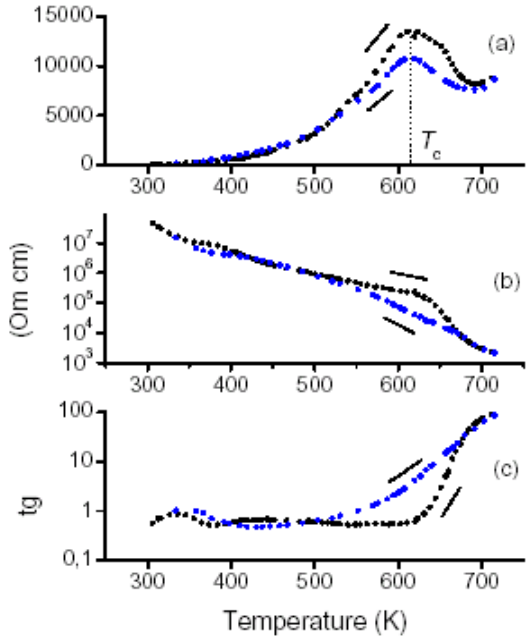

Fig. 4. Dependences of the dielectric permeability (a), specific resistance (b), and dielectric loss factor (c) vs. temperature for the CFO-PZT #3 structure.

It is seen in Fig.3 that for the sample NFO1-PZT #1 the dielectric constant has a value of $\varepsilon_{300}$ = 438 at room temperature $T$ = 300 K, reaches maximum value of $\varepsilon_{max} \sim$ 8880 at the temperature $T_c \approx$ 630 K, and than it decreases with further increase in the temperature. Maximum in the permittivity corresponds to a phase transition of PZT layers from the ferroelectric to the paraelectric state. Measured temperature of the phase transition $T_c$ and values of the permeabilities $\varepsilon_{300}$ and $\varepsilon_{max}$ are in a good agreement with data for bulk PZT samples [9]. The dependence $\varepsilon(T)$ corresponding to the sample cooling qualitatively traces similar dependence for the sample heating.

It follows from Fig. 3 that specific resistance of the sample $\rho$ = 6.7·$10^6$ Ω·cm at room temperature. As the sample is heated from room temperature to 470 K, the resistance remained nearly constant at $\sim 10^7$ Ω·cm. Then the resistance decreased nearly exponentially by $\sim$ 4 orders of magnitude at high temperatures. For sample cooling, the resistance was increased relative to sample heating and was restored to initial value at $\sim$ 400 K. Maximum difference in the resistances corresponding to heating and cooling of the sample is an order of magnitude in the vicinity of the phase transition temperature $T_c$.

Figure 3c shows the temperature dependence of the dielectric loss factor tgδ for the multilayer structure. The loss factor was decreased from ~0.7 to ~0.14 as the sample was heated from room temperature up to 470 K, and than it was increased exponentially at sample heating above 500 K. Increase in the loss factor with the temperature coincides with the sharp decrease in the specific resistance of the sample in the temperature range above 500 K. The hystersis in the temperature dependence of the loss factor is mainly due to dependence of the sample specific resistance on temperature.

Temperature dependences of the permittivity ε, specific resistance ρ, and the loss factor tgδ for the structures with cobalt ferrite, shown in Fig.4, are qualitatively similar to data in Fig.3. Note, that multilayer structures with nickel ferrite and structures with cobalt ferrite were sintered and slightly different temperatures. Difference in the preparation conditions, it seems, resulted in a small difference in the phase transition temperatures $T_c$ for the PZT layers and more significant difference in the values of dielectric constants.

### 4.2. Frequency dependence

Figure 5 shows typical dependence of ε, ρ, and tgδ for NFO2-PZT as a function of the frequency at room temperature. Note the logarithmic scale for frequency and resistance in Fig. 5b. It follows from Fig.5a that ε decreases by nearly an order in magnitude for all the structures as the frequency increases from 100 Hz to 10 kHz. The specific resistance for all the structures decreases by an order of magnitude as the frequency changes by two order in magnitude from 100 Hz to 10 kHz. Figure 5c demonstrates that the dielectric loss factor for all the structures ranges from 1 to 0.3 as the frequency of the modulation field changes by two orders in magnitude. For some structures the dielectric loss factor has a local maximum at $\sim$ 1 kHz. Similar temperature and frequency dependences of the dielectric parameters have been obtained for the multilayer structures containing cobalt ferrite, lithium ferrite and manganite films, described in Table 1. The dependences were qualitatively similar for all the structures..

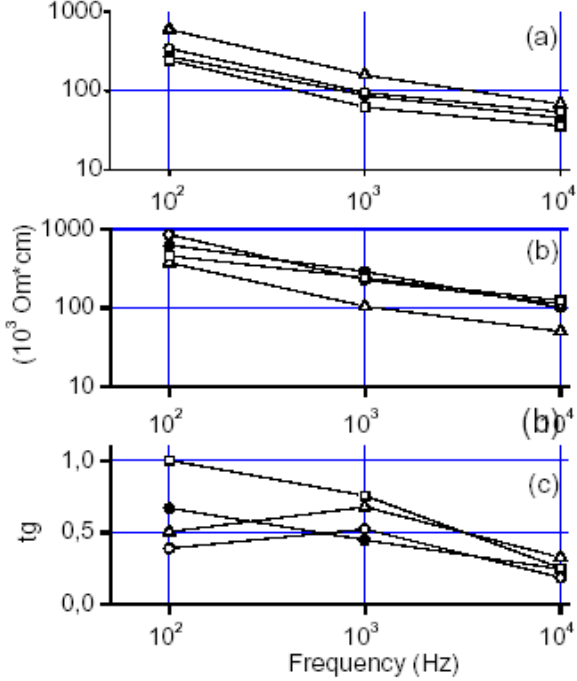

Fig. 5. Frequency dependences of the dielectric permeability (a), specific resistivity (b), and the dielectric loss factor (c) for the NFO2-PZT structures. Numbers near the curves correspond to numbers of the samples in Table 1. Samples notations: 1 –solid circle, 2 –open circle, 3- up triangle, 4- down triangle, 5 –square.

## 5. Model of multilayer structure and discussion

In order to explain the data in Fig.3-5, we consider a model of the multilayer structure consisting of a set of uniform layers of piezoelectric and ferromagnetic films connected in series. Let the piezoelectric has a dielectric permittivity $\varepsilon_p$ and a specific resistivity $\rho_p$, and the ferromagnet $\varepsilon_m$ and $\rho_m$. Then, one can show that effective values of dielectric constant $\varepsilon$, resistivity $\rho$, and loss factor tgδ are given by [13]:

$$\varepsilon^{-1} = p \cdot \varepsilon_p^{-1} + m \cdot \varepsilon_m^{-1} \qquad (1a)$$

$$\rho = p \cdot \rho_p + m \cdot \rho_m \qquad (1b)$$

$$tg\delta = \frac{p \cdot \varepsilon_p^{-1} + m \cdot \varepsilon_m^{-1}}{2\pi f \varepsilon_0 (p \cdot \rho_p + m \cdot \rho_m)}. \qquad (1c)$$

where $p = n \cdot d_p/d$ and $m = (n+1) \cdot d_m/d$ are the relative thickness of the piezoelectric and magnetic layers, respectively, and $d$ is the full thickness of the structure.

It follows from Eq.(1) that effective dielectric permittivity of the structure has to change from $\varepsilon_m$ to $\varepsilon_p$ and effective specific resistivity has to change from $\rho_m$ to $\rho_p$, as the relative thickness of PZT layers is varied in the limits $0 < p < 1$. Typical values of $\varepsilon_p \sim 10^3$ and $\rho_p \sim 10^9$ Ω·cm for PZT at room temperature are usually much higher than the corresponding values of $\varepsilon_m \sim 10^2$ and $\rho_m \sim 10^5$ Ω·cm for the ferrites [12]. Therefore, the values of the effective parameters for the multilayer structure have to be determined mainly by electrical properties of the PZT layers. In the structures investigated the relative thickness of the PZT was varied within the limits of 0.2 ≤ $p$ ≤ 0.8. This had to bring out a decrease of ε and ρ of no more than ~4 times in comparison with corresponding values for bulk PZT samples.

It follows from Fig. 3a, that the sample NFO1-PZT # 2 with $p$=0.66 does have the dielectric constant which in several times smaller than that one for bulk PZT. At the same time, the sample CFO-PZT # 3 with $p$=0.48 (see Fig. 4A) has a dielectric constant which is quite small. These observations point to the possible presence of "shorts" between ferrite layers in the structure (see Fig. 1). The "shorts" give rise to a "turning-off" of a number of PZT layers that result in a decrease in the measured dielectric constant. Possible presence of "shorts" in the structures is also confirmed by temperature dependences of specific resistivity ρ($T$), shown in Fig. 3b and 4b. The effective resistivity of the structures is ~ 3 orders of magnitude smaller than the expected value.

Appearance of a hysteresis in ρ($T$), it seems, is due to the hopping-type mechanism of conductivity in the nickel ferrite layers. The measurements carried out on individual films of nickel ferrite and PZT showed that hysteretic dependence was observed for the ferrite films and was not observed for the PZT films.

The model also gives qualitative explanation for frequency dependences of dielectric constant and the loss factor in Fig. 5. One can see that effective ε at all the frequencies decreases with a decrease in the relative content of PZT layers in the structures. Equation (1c) predicts a slow variation of the dielectric loss factor tgδ of the multilayer structure with frequency as seen in Fig.5.

## 6. Pyroelectric properties

### 6.1. Thermoelectric current and pyroelectric coefficient

The pyroelectric effect was investigated in all the structures in Table 1. In order to avoid irreversible depolarization of the structures, investigations of their characteristics were carried out within the temperature range of 290-400 K, that is at temperatures well below the phase transition temperature $T_c \sim$ 620 K for the PZT layers.

Fig. 6 shows typical results on measured dependence of the current density $j$ vs. temperature for NFO1-PZT # 2. The data are for increasing or decreasing temperatures. One observes a linear increase in the current with temperature up to ~320 K. The current remained nearly constant within the temperature range ~320-380 K. We notice an exponentially increasing current with further increase in T.

Upon cooling the sample, the current decreases exponentially and changes its sign at ~370 K. The current remained nearly constant with further cooling of the sample. In the temperature range 305 < $T$ < 370 K, the current direction is reversed. Thus one can identify three characteristics regions, "A", "B", and "C" in Fig. 6, that correspond to three different variations in the current. In the region A, $j$ vs T is linear. Within the temperature region "B" a pyroelectric current $j_P$ provides most of the total measured current. This pyroelectric current is created by polarized PZT layers as the temperature changes. Magnitude of the pyroelectric current is proportional to the rate of temperature variation and results in a constant current in this temperature region. The pyroelectric nature is confirmed by the positive and negative currents for heating and cooling, respectively.

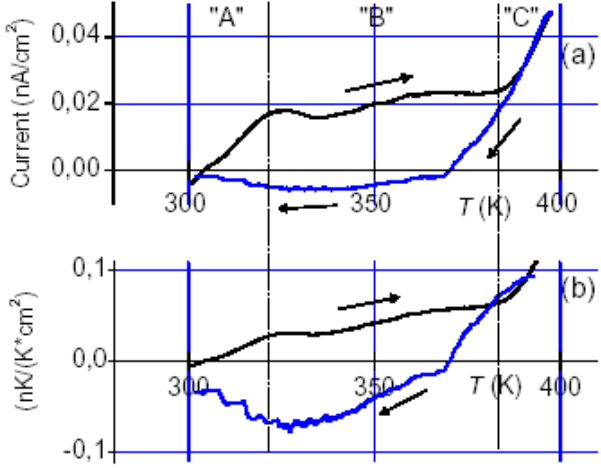

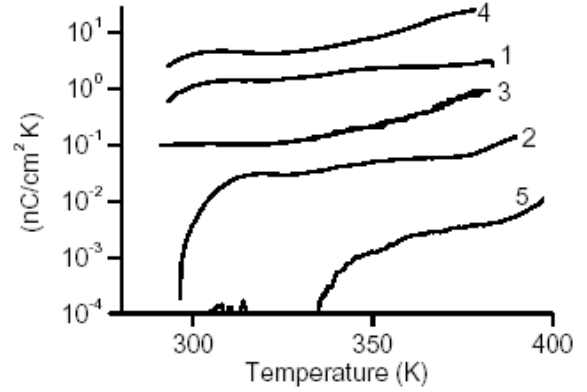

Fig. 6. Dependences of the current density (a) and coefficient γ(b) vs. temperature *T* for the NFO2-PZT # 2 structure. Temperature regions "A", "B" and "C" correspond to different physical mechanisms of the current formation.

Fig. 7. Temperature dependences of pyroelectric coefficient for the NFO1-PZT structures.

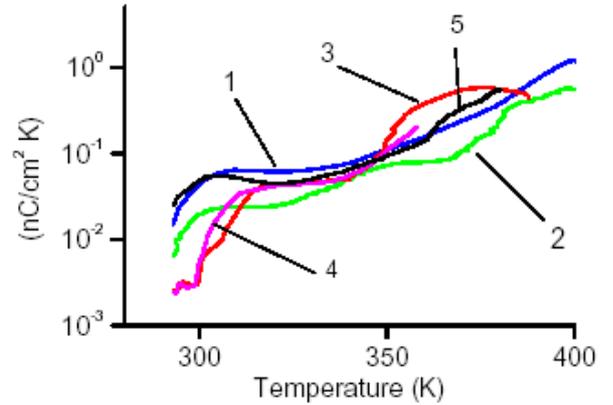

Fig. 8. Temperature dependences of pyroelectric coefficient for the NFO2-PZT structures.

Within the temperature region "C" the thermo-current $j_T$ plays the main role. The thermo-current arises because of redistribution of the charges in piezoelectric and magnetic layers of the structure. Magnitude of the thermo-current grows exponentially as the temperature of the sample increases. Magnitude and sign of the thermo-current depend on specific distribution of charges inside the structure layers. This current does not change sign when switched from heating to cooling of the sample.

Figure 6b shows dependence of the coefficient γ vs. temperature for heating and cooling for the sample NFO2-PZT # 2. The curves were calculated using the data of Fig. 6a and the formula γ = *j*/|*dT*/*dt*|. As it was mentioned above, the coefficient γ characterizes the pyroelectric effect only in the temperature region "B" (~305 – 370 K). Within this region γ has a negative sign for the sample cooling process because the current changes its direction. In the temperature region "C", the main contribution to the coefficient γ is due to the thermo-current, which is much higher than the pyroelectric current. This results in an exponential grows of the coefficient γ with temperature. Furthermore, to describe pyroelectric properties of the samples, we will use only the curve γ(*T*), corresponding to the sample heating.

### 6.2. Pyroelectric coefficient for different multilayer structures

Measurements of the current density as function of the temperature were carried out for all the structures in Table 1. Calculated temperature dependences of the coefficient γ for different structures are shown in Fig. 7-11. Numbers near the curves on the figures correspond to notations of the samples in Table 1.

Almost all of the multilayer structures show the pyroelectric effect due to nonzero dielectric polarization of the piezoelectric layers. But the magnitude of pyroelectric coefficient falls in a wide range. For the NFO1-PZT system, the coefficient is in the range $10^{-2}$ to 10 nC/(cm²·K). In most cases, the pyroelectric coefficient is much smaller than for bulk PZT ceramics. For PZT at room temperature the measured value of γ is 40 nC/(cm²·K). Both CFO-PZT and LFO-PZT show substantial thermo-currents at temperatures above 350 K. Negative value of γ for CFO-PZT #3 at ~ 400K is also due to thermo-currents.

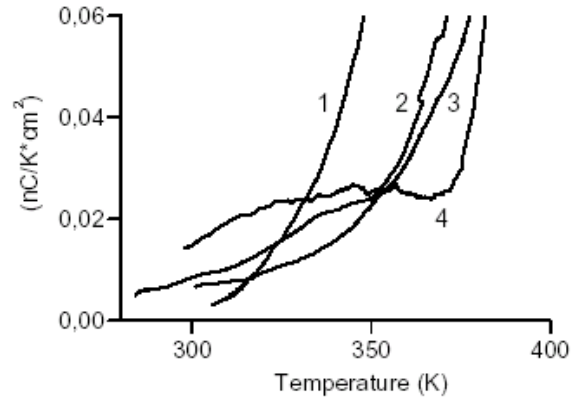

Fig. 10. Temperature dependences of pyroelectric coefficient for the LFO-PZT structures.

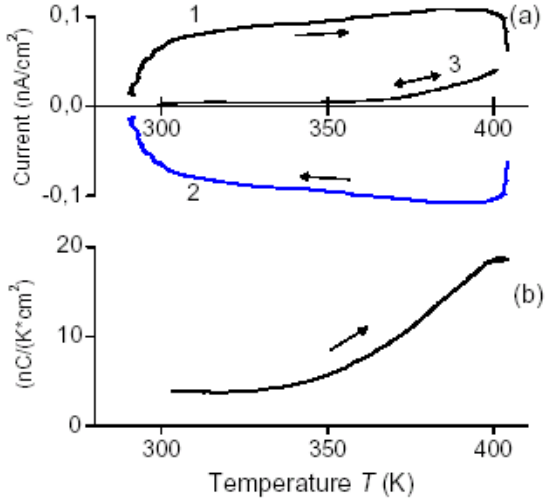

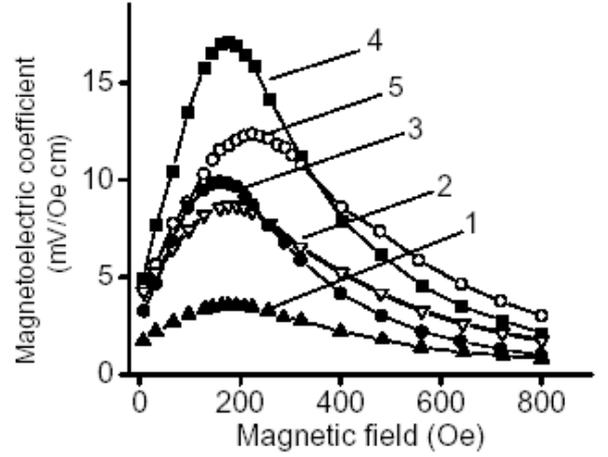

Fig. 12. Magnetoelectric coefficient $\alpha_E$ as a function of magnetic field $H$ for the NFO1-PZT structures measured at 400 Hz frequency.

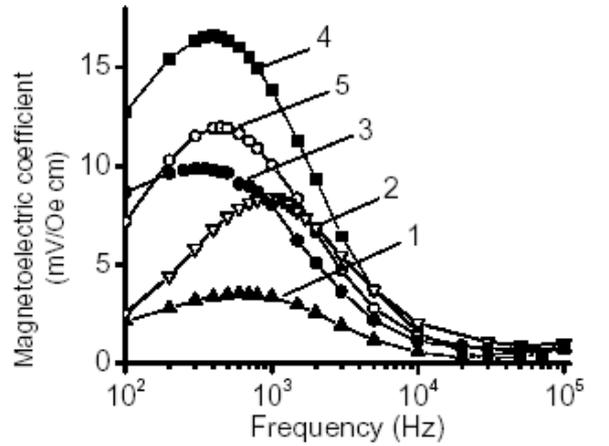

Fig. 13. Magnetoelectric coefficient $\alpha_E$ as a function of frequency $f$ for the NFO1-PZT structures measured for magnetic field of 200 Oe.

Fig. 11. Temperature dependences of the pyroelectric current for the LCM-PZT structure (curves 1 and 2) and the LSM-PZT structure (curve 3) ; b – temperature dependence of the $\gamma$ coefficient for the LCM-PZT structure corresponding the temperature increase.

The thermo-currents were not observed in LCM-PZT for the range room temperature to 400 K. This resulted in the formation of a very symmetrical dependence of the current vs. temperature sample heating and cooling cycles (see Fig.11). Such dependences are due to high degree of PZT polarization. But in LSMO-PZT, however, the thermo-current exceeded the pyroelectric current. The absence of pyroelectric effect is most likely due to low quality of the structure.

The results given above demonstrate significant disparity in the parameters characterizing the structures. This disparity is most likely due to defects in the structures. The main defect, it seems, is a formation of "shorts" between ferromagnetic or ferroelectric layers. The presence of even small number of such "shorts" results in considerable decrease in the transverse resistance of the structure, decreases efficiency of poling, and increases the thermo-currents. High porosity of the samples also leads to a weak pyroelectric and magnetoelectric effects.

## 7. Magnetoelectric effect

As mentioned in the Introduction, the magnetoelectric effect and the pyroelectric effect can coexist in multilayer ferrite-piezoelectric structures. A relation between magnetoelectric and pyroelectric coefficients can be obtained by comparing $\delta P$ caused by magnetic field $\delta H$ and temperature gradient $\delta T$:

$$\alpha_E \cdot \delta H = \gamma /(\varepsilon \varepsilon_0) \cdot \delta T . \qquad (2)$$

It follows from (2) that high pyroelectric coefficient and small dielectric permittivity are necessary to achieve high values of magnetoelectric coefficient.

In order to elucidate a relation between properties of multilayer structures and magnetoelectric effect, investigation of magnetoelectric effect was made for all the structures listed in Table 1. Measurements of magnetoelectric coefficient $\alpha_E$. have been carried out as described in detail in [1]. The sample was magnetized tangentially with an external dc magnetic field of $H$=0-2 kOe. A modulating magnetic field with the magnitude of $\delta H = 1$ Oe and the frequency of $f = 0.1 – 100$ kHz was applied parallel to the structure plane as well. An alternative voltage $\delta U$ of the same frequency was generated due to magnetoelectric coupling.

Figure 12 shows dependences of the magnetoelectric coefficient vs. magnetic field for the NFO2-PZT structures measured at $f = 400$ Hz. The coefficient reaches its maximum value $\alpha_E$ =18 mV/(Oe cm) for $H \sim 200$ Oe. The maximum in the $\alpha_E(H)$ curve corresponds to a point of maximum slope in the magnetostriction $\lambda$ vs. magnetic field dependence for the ferromagnetic layer [14]. Figure 13 shows dependences of magnetoelectric coefficient vs. modulation field frequency for the same NFO2-PZT structures measured at $H$=200 Oe. A decrease in the magnetoelectric coefficient in the low frequency region is probably due to a finite nonzero conductivity of the structure, which results in a relaxation of electrical charges generated on the sample surfaces. A drop in the magnetoelectric coefficient in the high frequency region is due to decrease in the specific conductivity of the layers and frequency dependence of the dielectric constant of the structures.

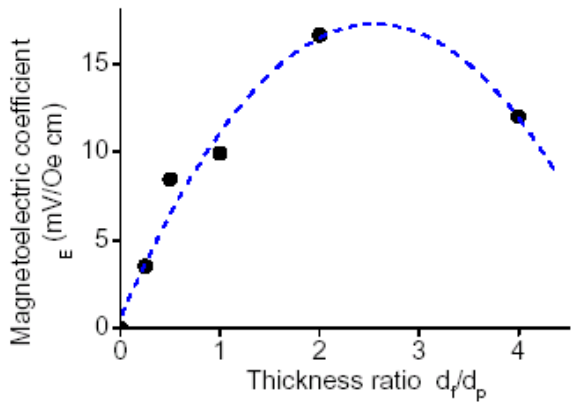

Fig. 14. Dependence of the magnetoelectric coefficient $\alpha_E$ vs. the ratio of the ferrite layer thickness to the piezoelectric layer thickness for the NFO1-PZT structure.

Figure 14 shows dependence of magnetoelectric coefficient as a function of relative concentration of PZT in the NFO1-PZT structures measured at magnetic field and frequency corresponding maximum values of the coefficient. Dashed lines in the figure are guide to the eyes. In accordance with theoretical predictions [15], the magnetoelectric coefficient reaches its maximum value as the nickel ferrite layer thickness is approximately double the PZT layer thickness.

A similar magnetic field and frequency dependences were measured for other structures listed in Table 1

## 8. Conclusion

Results of detailed dielectric and pyroelectric characterization of multilayer ferrite-PZT and manganite-PZT are provided here. The studies are important for information on the quality of the films. Multilayers with that show weak pyroelectric coupling are found to have low resistivity and small ME coefficients. The multilayers discussed here show an order of magnitude smaller ME coefficients compared to structurally superior samples.[1,14]

## Acknowledgements

The work was supported by a grant from the National Science Foundation (DMR-0302254).